\documentclass[aip, apl, amsmath, amssymb]{revtex4-1}
\usepackage{graphicx}
\usepackage{url}

\begin{document}
\bibliographystyle{aipnum4-1}

\title{Low Temperature Electrostatic Force Microscopy of a Deep Two Dimensional Electron Gas using a Quartz Tuning Fork}

\author{J.A. Hedberg}
\email[]{hedbergj@physics.mcgill.ca}
\author{A. Lal}
\author{Y. Miyahara}
\author{P. Gr\"utter}
\author{G. Gervais}
\author{M. Hilke}
\affiliation{Department of Physics, McGill University, Montreal, Canada, H3A 2T8}
\author{L. Pfeiffer}
\author{K.W. West}
\affiliation{Department of Electrical Engineering, Princeton University, Princeton, NJ, 08540 }

\date{17 August 2010}

\begin{abstract}
Using an ultra-low temperature, high magnetic field scanning probe microscope, we have measured electric potentials of a deeply buried two dimensional electron gas (2DEG). Relying on the capacitive coupling between the 2DEG and a resonant tip/cantilever structure, we can extract electrostatic potential information of the 2DEG from the dynamics of the oscillator. We present measurements using a quartz tuning fork oscillator and a 2DEG with a cleaved edge overgrowth structure. The sensitivity of the quartz tuning fork as force sensor is demonstrated by observation of Shubnikov de Haas oscillations at a large tip-2DEG separation distance of more than 500 nm.
\end{abstract}


\maketitle 

The simultaneous use of local probes and charge transport measurements offers exciting possibilities for exploring condensed matter systems. Traditionally, measuring the states of charge carriers in semiconductors involves creating electrical contacts to the structures which are fixed in position and must make physical and ohmic contact with the sample. Employing non-invasive scanning probe techniques in the study of electron transport phenomena has revealed information correlating spatial position with physical states. Several techniques exist that accomplish this: subsurface charge accumulation, \cite{Tessmer:1998p3987, Glicofridis:2002p5726} the related technique of scanning capacitance microscopy, \cite{Chakraborty:2004p5728} ac \cite{Woodside:2000p5731, McCormick:1998p5732} and dc\cite{Weitz:2000p1878} electrostatic force microscopy (EFM), and scanning electrometers using single electron transistors \cite{Zhitenev:2000p187}. Previous approaches combining scanned probes and a two dimensional electron gas (2DEG) focused on systems where the 2DEG was located close to the surface, usually residing at depths less than 100 nm from the surface. This work describes the development of ac-EFM as applied to a physical system where the 2DEG is buried much deeper, $\sim 500 \;\text{nm}$ below the surface. Deeply buried 2DEGs are less susceptible to scattering related mobility degradation, and are therefore more likely to exhibit exotic, high-mobility physics (Luttinger liquid behavior, fractional charges, Wigner Crystalization, e.g.) than shallower structures. 

\begin{figure}
\includegraphics[]{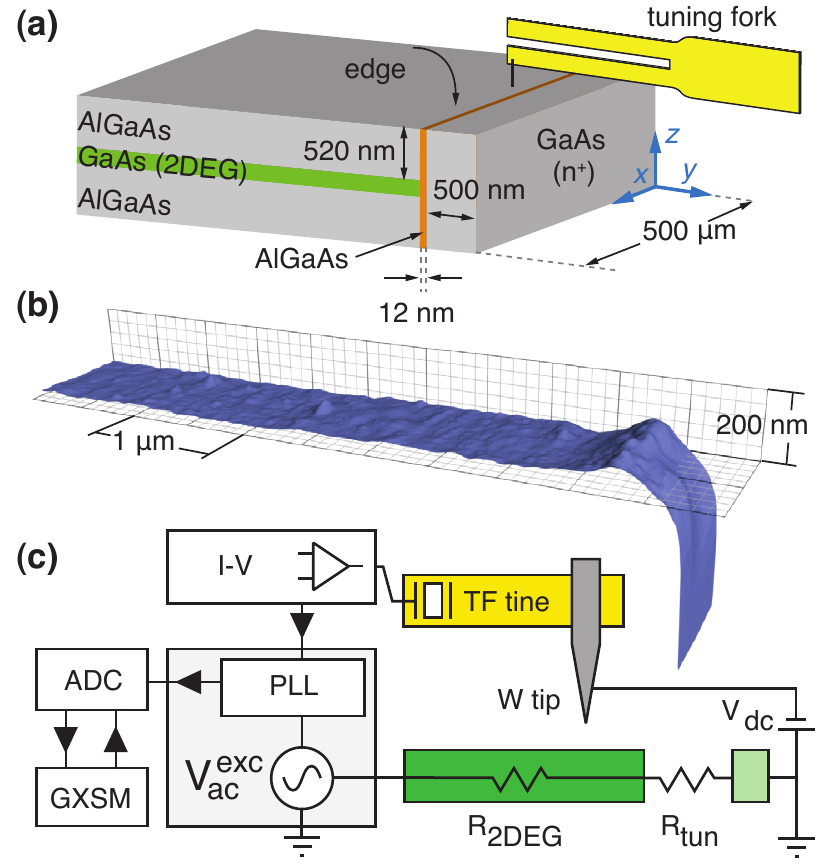}%
\caption{a) The relevant geometry of the cleaved edge overgrowth sample, b) Topographic image of the edge, and c) measurement schematic.}
\label{fig:scheme}%
\end{figure}

Our instrument consists of a homebuilt scanning probe microscope (SPM) mounted on the mixing chamber of a dilution refrigerator, capable of operation between 300 K and 300 mK. The microscope is housed within the bore of a superconducting magnet with a maximum field of 16 T. As a force sensor, we employ a piezoelectric quartz tuning fork (TF) in the Qplus configuration.\cite{Giessibl:2000p184} An etched tungsten wire, with a tip radius in the order of 10 nm is mounted to and electrically isolated from, the free tine of the tuning fork (resonant frequency, $f_0 \approx 20\;\text{kHz}$ and spring constant, $k \approx 1800 \;\text{N/m}$ ). Since the tungsten wire extends approximately 1 mm away from the TF, unwanted contributions from the plane of the TF electrodes are reduced as compared to conventional SPM cantilevers. The microscope is stable in high field ($16\; \text{T}$) environments, with negligible image distortion or translocation due to magnetic field influences. The quartz tuning fork has been shown to be capable of very small (sub \AA ) amplitudes of oscillation\cite{Gross:2009p5751}, and we have implemented sensitive electronics to measure oscillation amplitudes in the tens of picometers, while maintaining experimentally reasonable time scales. Similar instruments mentioned above rely on piezoresistive cantilevers as force sensors.\cite{Woodside:2000p5731, McCormick:1998p5732} Such sensors are much harder to use in dilution refrigerator based SPMs since their power dissipation during normal operation, ($\sim 1$ mW), would overwhelm the cooling power of the standard dilution unit, ($\sim 100\;\mu$W).\cite{Seo:2002p136}

The sample is a GaAs quantum well with a cleaved edge overgrowth structure\cite{Pfeiffer:1990p5386}. After the initial growth steps, the sample was cleaved, and a tunneling barrier and subsequent edge electrode were epitaxially grown on the resulting cleaved face. Measurements of the resistance of the tunneling barrier, $R_\text{tun}$, via the edge electrode can be used to characterize the edge states in exotic quantum hall systems with minimal intrusion.\cite{Chang:1996p5381,Hilke:2001p5376,Grayson:2006p5502}  Whereas previous studies involving buried 2DEGs and SPMs used relatively shallow electron gasses, \cite{Tessmer:1998p3987, Glicofridis:2002p5726, Chakraborty:2004p5728, Woodside:2000p5731, McCormick:1998p5732, Weitz:2000p1878, Zhitenev:2000p187, Baumgartner:2006p1557, Kicin:2004p56} the 2DEG in this sample is located 520 nm from the surface (See Figure \ref{fig:scheme}a for the relevant sample geometries). Indium contacts were diffused through the surface to create ohmic contact to the quantum well, thereby allowing for standard transport measurements as well.

To map the potential landscape of the buried 2DEG, an ac potential, $V^\text{exc}_\text{ac} $, was applied to the 2DEG at the resonance frequency, $f_0$, of the TF. Due to the capacitive coupling between the tip and the plane of the 2DEG, $C$, the electrostatic force on the TF is given by\cite{McCormick:1998p5732}:
\begin{equation} \label{eq:force}
F =\frac{1}{2} \frac{dC}{dz} \left( V_{\text{ac}} + V_{\text{dc}} + V_\text{cpd}\right)^2,
\end{equation}
where $z$ is the tip-sample separation, $V_{\text{ac}}$ is the local potential difference between the tip and sample at $f = f_0$, $V_{\text{dc}}$ is the local dc static potential, and  $V_\text{CPD}$ is the contact potential difference (CPD) arising from variations in the work function of the tip and sample materials. Expanding the potential term, and keeping only the terms that contribute to TF motion at $f_0$, we obtain for the excitation force: 
\begin{equation} \label{eq:forceAC}
F_\text{ac} = \frac{dC}{dz}V_\text{ac} (V_\text{dc} + V_\text{CPD}).
\end{equation}
Since, in the small amplitude limit, the deflection of the cantilever varies directly as the force, the oscillation amplitude can be used as a measure of the local potential, $V_\text{ac}$, provided there is little or no change in the tip-sample capacitance as a function of V. (Although the capacitance of a semi-conductor can be influenced by the tip-sample bias, we have seen no evidence of this effect in the current system, most likely a result of the large distance separating the tip and the 2DEG which reduces the charge accumulation and depletion effects of electrostatic fields associated with the tip.) Therefore, keeping $dC/dz$, $V_\text{dc}$, and $V_\text{CPD}$ constant, all other changes in the TF oscillation amplitude can be attributed to the deviations in the local potential. 

Normalization measurements were performed in which a uniform potential was applied to the entire sample, thus removing any contribution from the $dC/dz$ term as the probe approaches the edge. Also, measurements of the CPD showed no appreciable change as a function of location in the $y$ direction (See Fig. \ref{fig:scheme}a for sample axes). To record the dynamics of the TF, the current generated by its oscillatory motion was amplified with a custom I-V converter at room temperature, and fed to a Nanonis phase locked loop (PLL) which tracked the resonant frequency. Control of the scanning operation was handled by the GXSM software package in combination with the SignalRanger DSP card.\cite{Zahl:2003p72}(Fig. \ref{fig:scheme}c)

\begin{figure}
\includegraphics[]{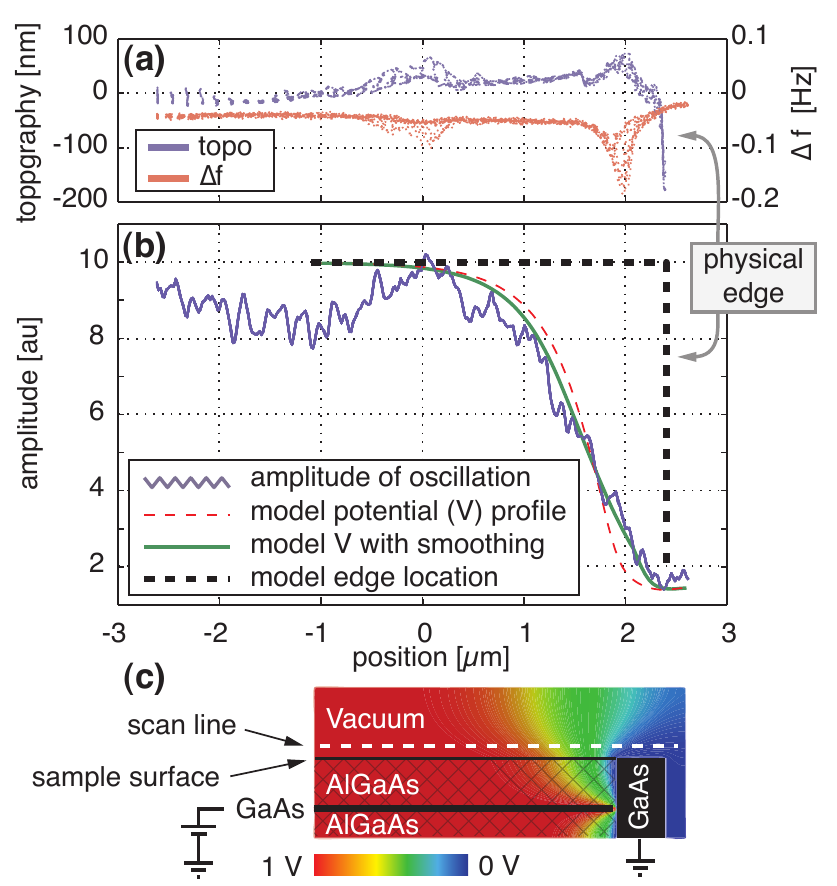}
\caption{Electrostatic model compared to measured TF oscillation amplitude. (a) shows cumulated cross sections of topographic and dc electrostatic images of the edge structure. The topographic images (topo) were acquired in NC-AFM mode and the electrostatic image ($\Delta f$) was obtained by scanning parallel to the surface at a lift height of 100 nm and applying a 10 V bias between the grounded sample and the tip. (b) The amplitude of the tuning fork (jagged blue) is shown plotted against the potential profiles estimated from finite element modeling of the structure. The dashed (red) line is taken directly from the model, and the solid (green) line is also from the model but smoothed to account for the finite size of the electrostatic tip radius.\cite{Sacha:2005p5819} (c) Electrostatic model of the cleaved edge overgrowth structure. The 2DEG was given a potential with respect to the grounded edge electrode. The dashed line shows the modeled scan line represented by the solid green line in (b). The scans were acquired at 400 mK in zero magnetic field.} 
\label{fig:edge}
\end{figure}

Exciting the cantilever via the described method, and monitoring the oscillation amplitude as the tip was scanned in a typical raster pattern towards the edge results in a 2D map of the potential. Since the resonant frequency, $f_0$, of the TF will also shift due to surface topography, we used the PLL to keep the excitation frequency, $f_{\text{exc}}$, matched to the changing $f_0$. Figure \ref{fig:edge}a,b shows three methods of visualizing the edge. Initially, a topographic image is acquired with a constant $\Delta f$ of 200 mHz, 0 V tip-sample bias, and oscillation amplitude of 10 nm, to verify the position of the tip with respect to the physical edge. Subsequently, the same scan area was repeated with a tip lift height of $\approx 100\; \text{nm}$ and a tip-sample dc bias of 10 V), while keeping the TF oscillating via mechanical excitation and the phase feedback loop engaged. This permits electrostatic force gradient microscopy of the edge.\cite{Seo:2002p136} Here, contrast from the topography remains visible and a plateau of constant frequency indicates the tip has moved past the edge, where $d^2C/dz^2$ is constant. These measurements are important as they confirm the location of the physical edge.

Eliminating the external mechanical excitation and exciting the TF solely via an ac potential on the 2DEG, we also recorded the TF oscillation amplitude as it varies with position. Keeping the edge electrode at 0V, the potential on the 2DEG was oscillated at $\approx 20$ kHz with an r.m.s. amplitude of 10 mV. A large dc bias of -25 V was also applied to the tip, to increase the oscillation amplitude, as expected from Eq. (\ref{eq:forceAC}). Fig. \ref{fig:edge}b shows a significant drop in the amplitude as the tip approached the edge. Of immediate interest is the length scale of the decline. The potential begins to drop approximately 2 $\mu$m before the physical edge. The shape and scale of this decline can readily be explained by electrostatic modeling of the edge structure and 2DEG geometry. A simple two conductor and dielectric model, (relative permittivity, $\epsilon_r\left(\text{AlGaAs}\right) = 13$) shown in Fig. \ref{fig:edge}c, offers a potential profile that fits well with that found from the amplitude of oscillation of the TF. (Note that although 1D line profiles are shown, the data was acquired with 2D area scans; no variation in the $x$ axis was observed.) 

\begin{figure}
\includegraphics[]{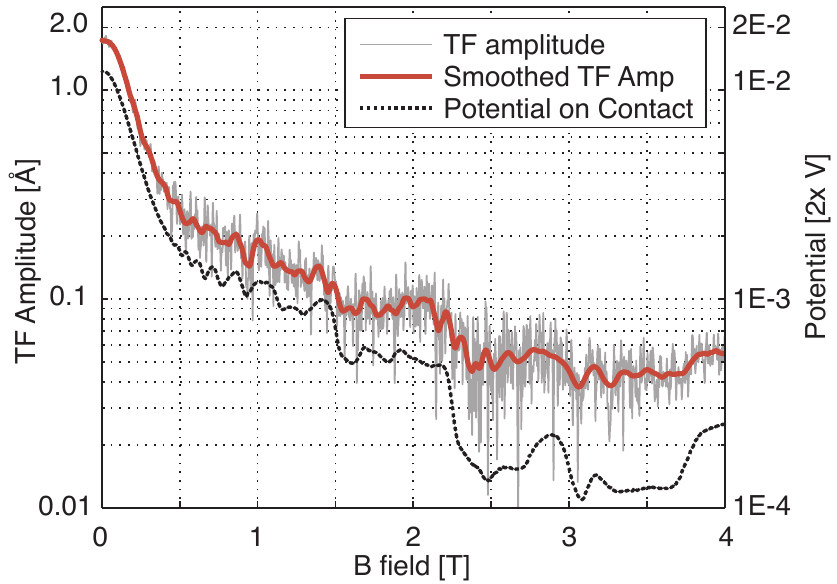}
\caption{Tuning fork amplitude showing 2DEG potential in a perpendicular B field. Since the force on the tuning fork was shown to be dependent on the potential in the 2DEG, we can use the oscillation amplitude to characterize the potentials created by the application of an external magnetic field. Shown in the solid line is a smoothed version of the raw tuning fork amplitude. The dashed line indicates the potential as measured via a contacted electrode. This data was acquired at 400 mK.}
\label{fig:Bfield}
\end{figure}

To further probe this interaction, we also applied a perpendicular magnetic field to elicit the 2D physics of the 2DEG. Shown in Fig. \ref{fig:Bfield} is the response of the tuning fork amplitude as a function of applied magnetic field. Also plotted is the potential measured on another indium contact on the 2DEG, near the tip position. The potential on this contact is measured via standard lock-in techniques using the ac excitation signal as the reference. The strong correlation between the TF amplitude and the directly measured potential confirms the original analysis from Eq. (\ref{eq:forceAC}), indicating the tip is acting as a potential probe. The potential measurement can also be converted to show the resistance of the 2DEG by considering the simple circuit model and relevant resistances shown in Fig. \ref{fig:scheme}c. Shubnikov de Haas oscillations are visible in the tuning fork amplitude above .5 T. From this we can extract a density of $\approx 1 \times 10^{11}\; \text{cm}^{-2}$.

In conclusion, we have demonstrated the ability to electro-mechanically couple a deeply buried 2DEG with a scannable tuning fork force sensor, at low temperatures and in high magnetic fields. This can allow for the extraction of electric potential information without introducing physically contacted electrodes. The high-sensitivity of the TF force sensor shows promise for the application of this technique to other physically interesting systems, where the region of interest is far from the probe, or in general when signals are small and close to noise limits.

We would also like to acknowledge V. Sazonova, C.R. Da Cunha and R. Bennewitz for contributions to the apparatus, as well as FQRNT, CIFAR, and NSERC for financial support.


\begin{thebibliography}{10}%
\makeatletter
\providecommand \@ifxundefined [1]{%
 \ifx #1\undefined \expandafter \@firstoftwo
 \else \expandafter \@secondoftwo
\fi
}%
\providecommand \@ifnum [1]{%
 \ifnum #1\expandafter \@firstoftwo
 \else \expandafter \@secondoftwo
\fi
}%
\providecommand \enquote [1]{``#1''}%
\providecommand \bibnamefont  [1]{#1}%
\providecommand \bibfnamefont [1]{#1}%
\providecommand \citenamefont [1]{#1}%
\providecommand\href[0]{\@sanitize\@href}%
\providecommand\@href[1]{\endgroup\@@startlink{#1}\endgroup\@@href}%
\providecommand\@@href[1]{#1\@@endlink}%
\providecommand \@sanitize [0]{\begingroup\catcode`\&12\catcode`\#12\relax}%
\@ifxundefined \pdfoutput {\@firstoftwo}{%
 \@ifnum{\z@=\pdfoutput}{\@firstoftwo}{\@secondoftwo}%
}{%
 \providecommand\@@startlink[1]{\leavevmode}%
 \providecommand\@@endlink[0]{}%
}{%
 \providecommand\@@startlink[1]{%
  \leavevmode
  \pdfstartlink
   attr{/Border[0 0 1 ]/H/I/C[0 1 1]}%
   user{/Subtype/Link/A<</Type/Action/S/URI/URI(#1)>>}%
  \relax
 }%
 \providecommand\@@endlink[0]{\pdfendlink}%
}%
\providecommand \url  [0]{\begingroup\@sanitize \@url }%
\providecommand \@url [1]{\endgroup\@href {#1}{\urlprefix}}%
\providecommand \urlprefix [0]{URL }%
\providecommand \Eprint[0]{\href }%
\@ifxundefined \urlstyle {%
  \providecommand \doi [1]{doi:\discretionary{}{}{}#1}%
}{%
  \providecommand \doi [0]{doi:\discretionary{}{}{}\begingroup
  \urlstyle{rm}\Url }%
}%
\providecommand \doibase [0]{http://dx.doi.org/}%
\providecommand \Doi[1]{\href{\doibase#1}}%
\providecommand \selectlanguage [0]{\@gobble}%
\providecommand \bibinfo [0]{\@secondoftwo}%
\providecommand \bibfield [0]{\@secondoftwo}%
\providecommand \translation [1]{[#1]}%
\providecommand \BibitemOpen[0]{}%
\providecommand \bibitemStop [0]{}%
\providecommand \bibitemNoStop [0]{.\EOS\space}%
\providecommand \EOS [0]{\spacefactor3000\relax}%
\providecommand \BibitemShut [1]{\csname bibitem#1\endcsname}%
\bibitem{Tessmer:1998p3987}%
  \BibitemOpen
  \bibfield{author}{%
  \bibinfo {author} {\bibfnamefont{S.}~\bibnamefont{Tessmer}}, \bibinfo
  {author} {\bibfnamefont{P.}~\bibnamefont{Glicofridis}}, \bibinfo {author}
  {\bibfnamefont{R.}~\bibnamefont{Ashoori}},\ and\ \bibinfo {author}
  {\bibfnamefont{L.}~\bibnamefont{Levitov}},\ }%
  \bibfield{journal}{%
  \bibinfo {journal} {Nature}\ }%
  \textbf{\bibinfo {volume} {392}},\ \bibinfo {pages} {51} (\bibinfo {year}
  {1998})\BibitemShut{NoStop}%
\bibitem{Glicofridis:2002p5726}%
  \BibitemOpen
  \bibfield{author}{%
  \bibinfo {author} {\bibfnamefont{P.}~\bibnamefont{Glicofridis}}, \bibinfo
  {author} {\bibfnamefont{G.}~\bibnamefont{Finkelstein}}, \bibinfo {author}
  {\bibfnamefont{R.}~\bibnamefont{Ashoori}},\ and\ \bibinfo {author}
  {\bibfnamefont{M.}~\bibnamefont{Shayegan}},\ }%
  \bibfield{journal}{%
  \bibinfo {journal} {Phys. Rev. B}\ }%
  \textbf{\bibinfo {volume} {65}},\ \bibinfo {pages} {121312} (\bibinfo {year}
  {2002})\BibitemShut{NoStop}%
\bibitem{Chakraborty:2004p5728}%
  \BibitemOpen
  \bibfield{author}{%
  \bibinfo {author} {\bibfnamefont{S.}~\bibnamefont{Chakraborty}}, \bibinfo
  {author} {\bibfnamefont{I.}~\bibnamefont{Maasilta}}, \bibinfo {author}
  {\bibfnamefont{S.}~\bibnamefont{Tessmer}},\ and\ \bibinfo {author}
  {\bibfnamefont{M.}~\bibnamefont{Melloch}},\ }%
  \bibfield{journal}{%
  \bibinfo {journal} {Phys. Rev. B}\ }%
  \textbf{\bibinfo {volume} {69}},\ \bibinfo {pages} {73308} (\bibinfo {year}
  {2004})\BibitemShut{NoStop}%
\bibitem{Woodside:2000p5731}%
  \BibitemOpen
  \bibfield{author}{%
  \bibinfo {author} {\bibfnamefont{M.}~\bibnamefont{Woodside}}, \bibinfo
  {author} {\bibfnamefont{C.}~\bibnamefont{Vale}}, \bibinfo {author}
  {\bibfnamefont{K.~L.}\ \bibnamefont{McCormick}}, \bibinfo {author}
  {\bibfnamefont{P.~L.}\ \bibnamefont{McEuen}}, \bibinfo {author}
  {\bibfnamefont{C.}~\bibnamefont{Kadow}}, \bibinfo {author}
  {\bibfnamefont{K.}~\bibnamefont{Maranowski}},\ and\ \bibinfo {author}
  {\bibfnamefont{A.~C.}\ \bibnamefont{Gossard}},\ }%
  \bibfield{journal}{%
  \bibinfo {journal} {Physica E}\ }%
  \textbf{\bibinfo {volume} {6}},\ \bibinfo {pages} {238} (\bibinfo {year}
  {2000})\BibitemShut{NoStop}%
\bibitem{McCormick:1998p5732}%
  \BibitemOpen
  \bibfield{author}{%
  \bibinfo {author} {\bibfnamefont{K.~L.}\ \bibnamefont{McCormick}}, \bibinfo
  {author} {\bibfnamefont{M.~T.}\ \bibnamefont{Woodside}}, \bibinfo {author}
  {\bibfnamefont{M.}~\bibnamefont{Huang}}, \bibinfo {author}
  {\bibfnamefont{P.~L.}\ \bibnamefont{McEuen}}, \bibinfo {author}
  {\bibfnamefont{C.}~\bibnamefont{Duruoz}},\ and\ \bibinfo {author}
  {\bibfnamefont{J.~S.}\ \bibnamefont{Harris Jr}},\ }%
  \bibfield{journal}{%
  \bibinfo {journal} {Physica B}\ }%
  \textbf{\bibinfo {volume} {249}},\ \bibinfo {pages} {79} (\bibinfo {year}
  {1998})\BibitemShut{NoStop}%
\bibitem{Weitz:2000p1878}%
  \BibitemOpen
  \bibfield{author}{%
  \bibinfo {author} {\bibfnamefont{P.}~\bibnamefont{Weitz}}, \bibinfo {author}
  {\bibfnamefont{E.}~\bibnamefont{Ahlswede}}, \bibinfo {author}
  {\bibfnamefont{J.}~\bibnamefont{Weis}}, \bibinfo {author}
  {\bibfnamefont{K.}~\bibnamefont{Klitzing}},\ and\ \bibinfo {author}
  {\bibfnamefont{K.}~\bibnamefont{Eberl}},\ }%
  \bibfield{journal}{%
  \bibinfo {journal} {Physica E}\ }%
  \textbf{\bibinfo {volume} {6}},\ \bibinfo {pages} {247} (\bibinfo {year}
  {2000})\BibitemShut{NoStop}%
\bibitem{Zhitenev:2000p187}%
  \BibitemOpen
  \bibfield{author}{%
  \bibinfo {author} {\bibfnamefont{N.}~\bibnamefont{Zhitenev}}, \bibinfo
  {author} {\bibfnamefont{T.}~\bibnamefont{Fulton}}, \bibinfo {author}
  {\bibfnamefont{A.}~\bibnamefont{Yacoby}}, \bibinfo {author}
  {\bibfnamefont{H.}~\bibnamefont{Hess}}, \bibinfo {author}
  {\bibfnamefont{L.~N.}\ \bibnamefont{Pfeiffer}},\ and\ \bibinfo {author}
  {\bibfnamefont{K.~W.}\ \bibnamefont{West}},\ }%
  \bibfield{journal}{%
  \bibinfo {journal} {Nature}\ }%
  \textbf{\bibinfo {volume} {404}},\ \bibinfo {pages} {473} (\bibinfo {year}
  {2000})\BibitemShut{NoStop}%
\bibitem{Giessibl:2000p184}%
  \BibitemOpen
  \bibfield{author}{%
  \bibinfo {author} {\bibfnamefont{F.}~\bibnamefont{Giessibl}},\ }%
  \bibfield{journal}{%
  \bibinfo {journal} {Appl. Phys. Lett.}\ }%
  \textbf{\bibinfo {volume} {76}},\ \bibinfo {pages} {1470} (\bibinfo {year}
  {2000})\BibitemShut{NoStop}%
\bibitem{Gross:2009p5751}%
  \BibitemOpen
  \bibfield{author}{%
  \bibinfo {author} {\bibfnamefont{L.}~\bibnamefont{Gross}}, \bibinfo {author}
  {\bibfnamefont{F.}~\bibnamefont{Mohn}}, \bibinfo {author}
  {\bibfnamefont{P.}~\bibnamefont{Liljeroth}}, \bibinfo {author}
  {\bibfnamefont{J.}~\bibnamefont{Repp}}, \bibinfo {author}
  {\bibfnamefont{F.}~\bibnamefont{Giessibl}},\ and\ \bibinfo {author}
  {\bibfnamefont{G.}~\bibnamefont{Meyer}},\ }%
  \bibfield{journal}{%
  \bibinfo {journal} {Science}\ }%
  \textbf{\bibinfo {volume} {324}},\ \bibinfo {pages} {1428} (\bibinfo {year}
  {2009})\BibitemShut{NoStop}%
 \bibitem{Seo:2002p136}%
  \BibitemOpen
  \bibfield{author}{%
  \bibinfo {author} {\bibfnamefont{Y.}~\bibnamefont{Seo}}, \bibinfo {author}
  {\bibfnamefont{W.}~\bibnamefont{Jhe}},\ and\ \bibinfo {author}
  {\bibfnamefont{C.}~\bibnamefont{Hwang}},\ }%
  \bibfield{journal}{%
  \bibinfo {journal} {Appl. Phys. Lett.}\ }%
  \textbf{\bibinfo {volume} {80}},\ \bibinfo {pages} {4324} (\bibinfo {year}
  {2002})\BibitemShut{NoStop}%
\bibitem{Pfeiffer:1990p5386}%
  \BibitemOpen
  \bibfield{author}{%
  \bibinfo {author} {\bibfnamefont{L.}~\bibnamefont{Pfeiffer}}, \bibinfo
  {author} {\bibfnamefont{K.}~\bibnamefont{West}}, \bibinfo {author}
  {\bibfnamefont{H.}~\bibnamefont{Stormer}}, \bibinfo {author}
  {\bibfnamefont{J.}~\bibnamefont{Eisenstein}}, \bibinfo {author}
  {\bibfnamefont{K.}~\bibnamefont{Baldwin}}, \bibinfo {author}
  {\bibfnamefont{D.}~\bibnamefont{Gershoni}},\ and\ \bibinfo {author}
  {\bibfnamefont{J.}~\bibnamefont{Spector}},\ }%
  \bibfield{journal}{%
  \bibinfo {journal} {Appl. Phys. Lett.}\ }%
  \textbf{\bibinfo {volume} {56}},\ \bibinfo {pages} {1697} (\bibinfo {year}
  {1990})\BibitemShut{NoStop}%
\bibitem{Chang:1996p5381}%
  \BibitemOpen
  \bibfield{author}{%
  \bibinfo {author} {\bibfnamefont{A.}~\bibnamefont{Chang}}, \bibinfo {author}
  {\bibfnamefont{L.}~\bibnamefont{Pfeiffer}},\ and\ \bibinfo {author}
  {\bibfnamefont{K.}~\bibnamefont{West}},\ }%
  \bibfield{journal}{%
  \bibinfo {journal} {Phys. Rev. Lett.}\ }%
  \textbf{\bibinfo {volume} {77}},\ \bibinfo {pages} {2538} (\bibinfo {year}
  {1996})\BibitemShut{NoStop}%
\bibitem{Hilke:2001p5376}%
  \BibitemOpen
  \bibfield{author}{%
  \bibinfo {author} {\bibfnamefont{M.}~\bibnamefont{Hilke}}, \bibinfo {author}
  {\bibfnamefont{D.}~\bibnamefont{Tsui}}, \bibinfo {author}
  {\bibfnamefont{M.}~\bibnamefont{Grayson}}, \bibinfo {author}
  {\bibfnamefont{L.}~\bibnamefont{Pfeiffer}},\ and\ \bibinfo {author}
  {\bibfnamefont{K.}~\bibnamefont{West}},\ }%
  \bibfield{journal}{%
  \bibinfo {journal} {Phys. Rev. Lett.}\ }%
  \textbf{\bibinfo {volume} {87}},\ \bibinfo {pages} {186806} (\bibinfo {year}
  {2001})\BibitemShut{NoStop}%
\bibitem{Grayson:2006p5502}%
  \BibitemOpen
  \bibfield{author}{%
  \bibinfo {author} {\bibfnamefont{M.}~\bibnamefont{Grayson}},\ }%
  \bibfield{journal}{%
  \bibinfo {journal} {Solid State Commun.}\ }%
  \textbf{\bibinfo {volume} {140}},\ \bibinfo {pages} {66} (\bibinfo {year}
  {2006})\BibitemShut{NoStop}%
\bibitem{Baumgartner:2006p1557}%
  \BibitemOpen
  \bibfield{author}{%
  \bibinfo {author} {\bibfnamefont{A.}~\bibnamefont{Baumgartner}}, \bibinfo
  {author} {\bibfnamefont{T.}~\bibnamefont{Ihn}}, \bibinfo {author}
  {\bibfnamefont{K.}~\bibnamefont{Ensslin}}, \bibinfo {author}
  {\bibfnamefont{G.}~\bibnamefont{Papp}}, \bibinfo {author}
  {\bibfnamefont{F.}~\bibnamefont{Peeters}}, \bibinfo {author}
  {\bibfnamefont{K.}~\bibnamefont{Maranowski}},\ and\ \bibinfo {author}
  {\bibfnamefont{A.}~\bibnamefont{Gossard}},\ }%
  \bibfield{journal}{%
  \bibinfo {journal} {Phys. Rev. B}\ }%
  \textbf{\bibinfo {volume} {74}},\ \bibinfo {pages} {165426} (\bibinfo {year}
  {2006})\BibitemShut{NoStop}%
\bibitem{Kicin:2004p56}%
  \BibitemOpen
  \bibfield{author}{%
  \bibinfo {author} {\bibfnamefont{S.}~\bibnamefont{Kicin}}, \bibinfo {author}
  {\bibfnamefont{A.}~\bibnamefont{Pioda}}, \bibinfo {author}
  {\bibfnamefont{T.}~\bibnamefont{Ihn}}, \bibinfo {author}
  {\bibfnamefont{K.}~\bibnamefont{Ensslin}}, \bibinfo {author}
  {\bibfnamefont{D.}~\bibnamefont{Driscoll}},\ and\ \bibinfo {author}
  {\bibfnamefont{A.}~\bibnamefont{Gossard}},\ }%
  \bibfield{journal}{%
  \bibinfo {journal} {Physica E}\ }%
  \textbf{\bibinfo {volume} {21}},\ \bibinfo {pages} {708} (\bibinfo {year}
  {2004})\BibitemShut{NoStop}%
\bibitem{Zahl:2003p72}%
  \BibitemOpen
  \bibfield{author}{%
  \bibinfo {author} {\bibfnamefont{P.}~\bibnamefont{Zahl}}, \bibinfo {author}
  {\bibfnamefont{M.}~\bibnamefont{Bierkandt}}, \bibinfo {author}
  {\bibfnamefont{S.}~\bibnamefont{Schr{\"o}der}},\ and\ \bibinfo {author}
  {\bibfnamefont{A.}~\bibnamefont{Klust}},\ }%
  \bibfield{journal}{%
  \bibinfo {journal} {Rev. Sci. Instrum.}\ }%
  \textbf{\bibinfo {volume} {74}},\ \bibinfo {pages} {1222} (\bibinfo {year}
  {2003})\BibitemShut{NoStop}%
\bibitem{Sacha:2005p5819}%
  \BibitemOpen
  \bibfield{author}{%
  \bibinfo {author} {\bibfnamefont{G.}~\bibnamefont{Sacha}}, \bibinfo {author}
  {\bibfnamefont{A.}~\bibnamefont{Verdaguer}}, \bibinfo {author}
  {\bibfnamefont{J.}~\bibnamefont{Martinez}}, \bibinfo {author}
  {\bibfnamefont{J.}~\bibnamefont{S{\'a}enz}}, \bibinfo {author}
  {\bibfnamefont{D.}~\bibnamefont{Ogletree}},\ and\ \bibinfo {author}
  {\bibfnamefont{M.}~\bibnamefont{Salmeron}},\ }%
  \bibfield{journal}{%
  \bibinfo {journal} {Appl. Phys. Lett.}\ }%
  \textbf{\bibinfo {volume} {86}},\ \bibinfo {pages} {123101} (\bibinfo {year}
  {2005})\BibitemShut{NoStop}%

\end{thebibliography}
\end{document}